\documentclass[english,twocolumn,floatfix,prb,aps,showpacs,bibliography]{revtex4-1}
\usepackage[T1]{fontenc}
\usepackage[latin9]{inputenc}
\usepackage{babel}
\usepackage{amsmath}
\usepackage{amssymb}
\usepackage[unicode=true,pdfusetitle,
 bookmarks=true,bookmarksnumbered=false,bookmarksopen=false,
 breaklinks=false,pdfborder={0 0 1},backref=false,colorlinks=false]
 {hyperref}

\makeatletter
\usepackage{graphicx,color,curves,epic}

\makeatother

\begin{document}
\global\long\def\i{\imath}

\global\long\def\plll{P_{\text{\ensuremath{\mathrm{LLL}}}}}

\global\long\def\S{\mathcal{S}}

\global\long\def\T{\mathcal{T}}

\global\long\def\elliptic#1#2#3{\vartheta_{#1}\left(#2\middle|#3\right)}

\global\long\def\ket#1{\left|#1\right\rangle }

\global\long\def\bra#1{\left\langle #1\right|}

\global\long\def\braket#1#2{\left\langle #1\middle|#2\right\rangle }

\global\long\def\ketbra#1#2{\left|#1\vphantom{#2}\right\rangle \left\langle \vphantom{#1}#2\right|}

\global\long\def\braOket#1#2#3{\left\langle #1\middle|#2\middle|#3\right\rangle }

\global\long\def\weierstrauss#1#2#3{\sigma_{#2}^{\left(#1\right)}\left(#3\right)}

\newcommand{\FIXME}[1]{({\bf FIXME: #1})}

\title{On the modular covariance properties of composite fermions on the
torus}

\author{Mikael Fremling$^{1,2}$}

\affiliation{$^{1}$Institute for Theoretical Physics, Center for Extreme Matter
and Emergent Phenomena, Utrecht University, Princetonplein 5, 3584
CC Utrecht, the Netherlands}

\affiliation{$^{2}$Department of Theoretical Physics, Maynooth University, Maynooth,
Co. Kildare, W23 HW31, Ireland}
\begin{abstract}
In this work we show that the composite fermion construction for the
torus geometry is modular covariant. We show that this is the case
both before and after projection, and that modular covariance properties
are preserved under both exact projection and under JK projection
which was recently introduced by Pu, Wu, and Jain (PRB 96, 195302
(2017)). It is crucial for the modular properties to hold that the
CF state is a proper state, \emph{i.e. }that there are no holes in
the occupied $\Lambda$-levels.
\end{abstract}
\maketitle

\section{Introduction}

In the study of the fractional quantum hall effect, a prominent role
has been played by the construction of trial wave functions, dating
back to Laughlin's wave function\cite{Laughlin1983a} more than three
decades ago. The first formulation, which was for a finite quantum
liquid on an infinite plane, was subsequently generalized to both
a sphere\cite{Haldane1983} and torus\cite{Haldane1985b}. In the
following years other trial wave functions where also constructed
on the three geometries, such as the Pfaffian\cite{Moore1991,Morf1998,Rezayi2000}.
These three geometries: plane, sphere and torus, have since then been
the canonical playground for fractional quantum hall trial wave functions.

In this work, we build upon and extend recent developments\cite{Pu2017}
in constructing Jain-Kamilla projected wave functions for composite
fermions on the torus geometry. Composite fermions (CF), are straight
forward to write down in unprojected form\cite{Jain1989,Dev1992,Hermanns2013}
on all the three above mentioned geometries. To obtain physical wave
functions\emph{, i.e.} that reside in the lowest Landau level (LLL),
the CF wave functions however need to be projected onto that LLL.
This can be achieved either analytically\cite{Girvin1984b} or via
the Jain-Kamilla (JK) projection\cite{Jain1997}. The former is exact,
but numerically inefficient, and the latter is an uncontrolled approximation,
but numerically fast. It was early understood how to perform both
of these projections on the plane and sphere, but the torus geometry
proved more difficult, mainly due to technical difficulties with the
non-trivial interplay of boundary conditions and the action of derivatives
on quasi-periodic wave functions. The first successful attempts in
this direction was taken by Ref.\,\onlinecite{Hermanns2013} for the
analytical projection. 

In a parallel development, trial wave functions on the torus was also
developed for the Jain series with the help of CFT\cite{Hermanns2008,Fremling2014a,Hansson2017}
techniques. Recently DMRG methods have also been extended to the cylinder
geometry for the Laughlin state\cite{Zaletel2012}, it's quasi-particles\cite{Kjall2018},
and states higher up in the Hierarchy\cite{Zaletel2015}. See also
the construction of quasi-particles for the Laughlin state on the
Torus\cite{Greiter2016}.

Recently however, Pu, Wu and Jain\cite{Pu2017} (PWJ) managed to extend
the JK projection-scheme to also encompass the torus. The same techniques
where later used in Ref.\,\onlinecite{Pu2018} to study the composite
fermion Fermi liquid, which had previously been examined by other
numerical techniques\cite{Rezayi2000,Wang2017b,Fremling2018,Geraedts2018}.

In comparison to the other two geometries, the torus comes with an
extra parameter $\tau$, which controls its geometry. The parameter
$\tau$ is important since multiple values of $\tau$ may correspond
to the same physical geometry. This redundancy poses additional physical
constraints on the trial wave functions that are not present on the
plane and sphere, where it's sufficient to respect the boundary conditions.
It is therefore of great importance that wave functions defined on
the torus, not only have correct boundary conditions, but also that
wave functions at different (but physically equivalent) $\tau$ span
the same space of wave functions. The mapping from one value of $\tau$
to a physically equivalent value is a modular transformation and comes
in two flavors; the $T$-transform which sends $\tau\to\tau+1$ and
the $\S$-transform which sends $\tau\to-\frac{1}{\tau}$. The former
is a remapping of the torus lattice vectors, and the latter is a rotation
that interchanges the order of the vectors. Sets of wave functions
that span the same physical space before and after a the above mentioned
modular transformations have the property of modular covariance. The
modular covariance property was of great importance to compute e.g
Hall viscosity\cite{Read2009a,Read2008,Read2011,Fremling2014a}.

The property of modular covariance is not guaranteed simply because
appropriate boundary conditions are imposed. This was made clear in
Ref.\,\onlinecite{Fremling2014a}, where it was shown that the primary
CFT correlation functions used to construct hierarchy wave functions
have the correct modular properties, but that the naive introduction
of a regularized derivative (as was previously done in Ref.\,\onlinecite{Hermanns2008})
broke modular covariance. The authors could find another regularization
which restored the modular covariance and as a positive side effect
also significantly improved the overlap with the coulomb ground state.

The property of modular covariance has never been proven for the composite
fermion states, neither before nor after projection, and that is what
we will do in this paper. On the route there we will also present
some (hopefully) useful reformulations and results of the PWJ approach.
The paper is organized as follows: In Section (\ref{sec:Torus_and_CF})
we introduce the torus geometry and the single particle wave functions.
In Section \ref{sec:GJ-in-tau} we discuss the Girvin-Jach rule in
$\tau$-gauge. In Section \ref{sec:CFs_on_torus} we briefly discuss
the CF construction on the torus and in Section (\ref{sec:Modified-JK-projection})
we discuss the modification of the Grivin-Jach rule that is necessary
to obtain periodic boundary conditions for the Jain-Kamilla projection.
In section (\ref{sec:Modular-Covariance}) we derive the covariance
properties for unprotected as well as exactly projected and PWJ projected
CF:s and show that they all satisfy modular covariance. We end with
a discussion and outlook in Section (\ref{sec:Discussion}). Detailed
derivations are deferred to the Appendices.

\begin{figure}
\begin{centering}
\setlength{\unitlength}{1.5cm}
\begin{picture}(4,3.3)(0,-1)
\put(1.2,1.2){\large{$\tau=\frac{L_2}{L_1}=\tau_1+\i\tau_2$}}
\put(1.2,0.1){\Large{$L_1=L$}}
\put(-0.6,1){\Large{$L\tau_2$}}
\put(.2,1.65){\Large{$L\tau_1$}}
\put(0,0){\line(1,0){3}}
\put(0,0){\line(1,2){1}}
\put(3,0){\line(1,2){1}}
\put(1,2){\line(1,0){3}}
\dashline{0.2}(0,0)(0,2)
\dashline{0.2}(0,2)(1,2)
\put(0,0){\vector(2,-1){1.5}}
\put(1.2,-.5){\large{$\vec{\tilde A} 
= \frac{\tilde{y}B}{\tau_{2}} \left(\tau_{2},-\tau_{1}\right)$}}
\put(3.6,.2){\vector(1,0){.5}}
\put(3.6,.2){\vector(0,1){.5}}
\put(4.2,.1){$\tilde x$}
\put(3.6,.8){$\tilde y$}
\put(0.2,0.1){\vector(1,0){.5}}
\put(0.2,0.1){\vector(1,2){.333}}
\put(0.8,0.05){$x$}
\put(0.6,0.7){$y$}
\end{picture}
\par\end{centering}
\caption{The relationship between the Cartesian coordinates $\left(\tilde{x},\tilde{y}\right)$
and the dimensionless coordinates $\left(x,y\right)$. In the figure
one can also see that $\tau_{1}$ is interpreted as the skewness,
and $\tau_{2}$ as the aspect ratio, of the torus. Note how the $\tau$-gauge
vector potential $\vec{\tilde{A}}$ is perpendicular to the vector
$\vec{\tau}=\left(\tau_{1},\tau_{2}\right)$. The area of the torus
is fixed to be $L^{2}\tau_{2}=2\pi N_{\phi}\ell_{B}$.\label{fig:Geometry}}
\end{figure}
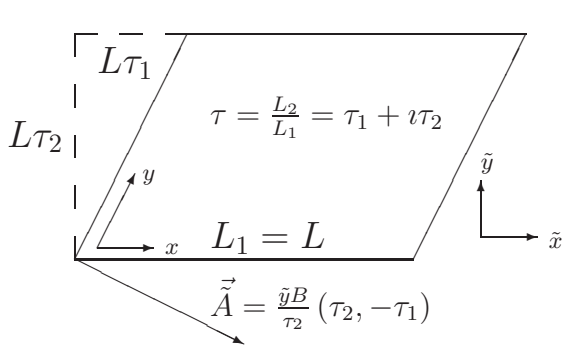

\section{The Torus and its wave functions\label{sec:Torus_and_CF}}

In this section we give a short recapitulation regarding the torus.
This also serves to define the notation that is used later in the
paper. The torus is defined by two axes $\boldsymbol{L}_{1}$, $\boldsymbol{L}_{2}$
on the plane and we will adopt the conventions that in complex coordinates
the axes are $L_{1}=L$ and $L_{2}=\tau L$, where $\tau=\tau_{1}+\i\tau_{2}$.
For coordinates we use the (unusual) convention that $z=\tilde{x}+\i\tilde{y}=L\left(x+\tau y\right)$
where $\tilde{x},\tilde{y}$ are the physical euclidean (dimensional)
coordinates and $x,y$ are the reduced (dimensionless) coordinates.
The reduced coordinates $x,y\in\left[0,1\right]$, defined on the
unit square are convenient since $x=1$ $\left(y=1\right)$ corresponds
to $z=L_{1}$ $\left(z=L_{2}\right)$. The two torus axes span an
area $\left|\boldsymbol{L}_{1}\times\boldsymbol{L}_{2}\right|=\tau_{2}L^{2}=2\pi N_{\phi}\ell^{2}$
where $\ell=\sqrt{\frac{\hbar e}{B}}$ is the magnetic length and
$N_{\phi}$ is the number of magnetic fluxes that penetrate it's surface.
See Fig. \ref{fig:Geometry} for an illustration of the coordinates
and gauge choice.

The single particle Hamiltonian is 
\begin{equation}
H=\sum_{i=\tilde{x},\tilde{y}}\frac{1}{2m}\left(p_{j}-e\tilde{A}_{j}\right)^{2},\label{eq:Hamiltonian}
\end{equation}
 where $p_{j}=\i\hbar\partial_{j}$ and $\vec{\tilde{A}}=\sum_{i=\tilde{x},\tilde{y}}\tilde{A}_{i}\hat{i}$
is a vector potential satisfying $\nabla\times\vec{\tilde{A}}=B\hat{z}$.
We will choose to work in the $\tau$-gauge, where the vector potential
is $\vec{\tilde{A}}=\frac{\tilde{y}B}{\tau_{2}}\left(\tau_{2},-\tau_{1}\right)$,
which is perpendicular to the vector $\vec{\tau}$. In reduced coordinates
the vector potential simplifies to $\vec{A}=\left(2\pi N_{\phi}yB,0\right)$
which is explicitly $\tau$-independent. In this work we will work
exclusively in $\tau$-gauge, as it is especially convenient to handle
modular transformations and boundary conditions at arbitrary $\tau$.
The Hamiltonian in (\ref{eq:Hamiltonian}) can be diagonalized by
introducing ladder operators, yielding the form $H=\hbar\omega_{B}\left(a^{\dagger}a+\frac{1}{2}\right)$,
where $\omega_{B}=\frac{eB}{m}$. The ladder operators in $\tau$-gauge
are
\begin{align}
a_{\tau} & =\sqrt{2}\left(\frac{\tau}{2}\frac{L}{\ell}y+\ell\partial_{\bar{z}}\right)\nonumber \\
a_{\tau}^{\dagger} & =\sqrt{2}\left(\frac{\bar{\tau}}{2}\frac{L}{\ell}y-\ell\partial_{z}\right),\label{eq:Ladder_tau}
\end{align}
 and satisfy $\left[a_{\tau}^{\dagger},a_{\tau}\right]=1$. Physical
wave functions $\psi\left(z\right)$, are quasi-periodic and obey
the boundary conditions 
\begin{equation}
\psi\left(z+L_{j}\right)=e^{\lambda_{j}\left(z,\bar{z}\right)}\psi\left(z\right)\label{eq:PBC}
\end{equation}
 where $\lambda_{j}\left(z,\bar{z}\right)$ depends on the gauge choice
$\vec{A}$. For $\tau$-gauge this is $\lambda_{j}=\delta_{j,2}2\pi N_{\phi}x$.

In $\tau$\textendash gauge, the shift operator and magnetic translation
operators are 
\begin{eqnarray}
\tilde{t}\left(\alpha L+\beta\tau L\right) & = & e^{\alpha\partial_{x}+\beta\partial_{y}}\label{eq:magnetic_translation}\\
t\left(\alpha L+\beta\tau L\right) & = & e^{\alpha\partial_{x}+\beta\partial_{y}+\i2\pi\beta N_{\phi}x}\nonumber 
\end{eqnarray}
 where it is the latter that defines periodic boundary conditions
$t\left(L_{j}\right)\psi=e^{\i\phi_{j}}\psi$. Note how we differentiate
between the shift operator $\tilde{t}\left(\tau L\right)$ and the
full magnetic translation operator $t\left(\tau L\right)$ such that
$t\left(\tau L\right)=e^{\i2\pi N_{\phi}x}\tilde{t}\left(\tau L\right)$.
General LLL wave functions in $\tau$-gauge take the form $\psi^{\left(A\right)}\left(z\right)=e^{\i2\pi Ay^{2}}f\left(z\right)$
where $f\left(z\right)$ is a holomorphic function and $A$ counts
the number of magnetic fluxes through the torus. The above formula
is particularity useful since if $\psi^{\left(A\right)}$ and $\psi^{\left(B\right)}$are
wave functions with boundary conditions $\phi_{A}$ and $\phi_{B}$
then $\psi^{\left(A+B\right)}=\psi^{\left(A\right)}\cdot\psi^{\left(B\right)}$
is automatically a wave function with boundary conditions $\phi_{A}+\phi_{B}$. 

The operator $t\left(\frac{L}{N_{\phi}}\right)$ commutes with the
operator $t\left(\tau L\right)$, and consequently can be used to
defined a basis of $N_{\phi}$ linearly independent states. The single
particle orbitals in the lowest Landau level \textendash{} in a basis
that diagonalizes $t\left(\frac{L}{N_{\phi}}\right)$ \textendash{}
can be written as 

\begin{eqnarray}
\phi_{i}^{\left(N_{\phi}\right)} & = & \frac{1}{\sqrt{\ell L\sqrt{\pi}}}e^{\i\pi\tau N_{\phi}y^{2}}\elliptic{\frac{i}{N_{\phi}},0}{\frac{N_{\phi}z}{L}}{N_{\phi}\tau}\nonumber \\
 & = & \frac{1}{\sqrt{\ell L\sqrt{\pi}}}\sum_{k\in\mathbb{Z}+\frac{i}{N_{\phi}}}e^{\i\pi N_{\phi}\tau\left(y+k\right)^{2}}e^{\i2\pi N_{\phi}kx}.\label{eq:LLL-wfn}
\end{eqnarray}
The function $\elliptic{a,b}z{\tau}=\sum_{k\in\mathbb{Z}+a}e^{\i\pi\tau k^{2}}e^{\i2\pi k\left(z+b\right)}$
is a generalized Jacobi theta function. The orbitals for the higher
landau levels are obtained by application of the raising operators
as $\phi_{j,n}=\frac{\left(a^{\dagger}\right)^{n}}{\sqrt{n!}}\phi_{j}$.
The explicit expression for the $n$:th Landau level orbitals, also
as eigenstates of $t\left(\frac{L}{N_{\phi}}\right)$, are 

\begin{eqnarray}
 &  & \phi_{i,n}^{\left(N_{\phi}\right)}=\mathcal{N}_{n}\sum_{k\in\mathbb{Z}+\frac{i}{N_{\phi}}}e^{\i\pi N_{\phi}\tau\left(y+k\right)^{2}}e^{\i2\pi N_{\phi}kx}H_{n}\left(\frac{\tau_{2}L}{\ell^{2}}\left(y+k\right)\right)\nonumber \\
 &  & =\mathcal{N}_{n}e^{\i\pi N_{\phi}\tau y^{2}}\sum_{k\in\mathbb{Z}+\frac{i}{N_{\phi}}}e^{\i\pi N_{\phi}\tau k^{2}}e^{\i2\pi N_{\phi}k\frac{z}{L}}H_{n}\left(\tilde{y}+\tau_{2}Lk\right),\label{eq:nLL-wfn}
\end{eqnarray}
 where $H_{n}$ is a Hermite polynomial and $\mathcal{N}_{n}=\frac{1}{\sqrt{2^{n}n!L\sqrt{\pi}}}$.
Note the appearance of the physical $\tilde{y}=L\tau y$ in the argument
of the Hermite polynomial. We will refer to $f_{i,n}^{\left(N_{\phi}\right)}$
as the holomorphic polynomial in $\phi_{i,n}^{\left(N_{\phi}\right)}=e^{\i\pi N_{\phi}\tau y^{2}}f_{i,n}^{\left(N_{\phi}\right)}$,
and we will often drop the momentum index $i$ for brevity.

In recent papers\cite{Wang2017b,Geraedts2018,Haldane2018,Haldane2018a},
Haldane has been advocating the use of Weierstrass $\sigma$-functions
over the traditionally used $\vartheta$-functions. In this paper
we follow in that tradition and define a generalized $\sigma$-function
in $\tau$-gauge as 

\begin{equation}
\sigma_{a,b}^{\left(n\right)}\left(z\right)=e^{\i\pi\tau ny^{2}}\elliptic{a,b}{\frac{nz}{L}}{n\tau}.\label{eq:sigma-function}
\end{equation}
 Comparing with (\ref{eq:LLL-wfn}) we have \emph{e.g.} that $\sigma_{\frac{i}{N_{\phi}},0}^{\left(N_{\phi}\right)}=\sqrt{\ell L\sqrt{\pi}}\phi_{i}^{\left(N_{\phi}\right)}$.
The Weierstrass functions builds in the quasi-period boundary conditions
and thus transform under coordinate changes as 
\begin{eqnarray*}
\sigma_{a,b}^{\left(n\right)}\left(z+L\right) & = & e^{\i2\pi an}\sigma_{a,b}^{\left(n\right)}\left(z\right)\\
\sigma_{a,b}^{\left(n\right)}\left(z+L\tau\right) & = & e^{-\i2\pi\left(nx+b\right)}\sigma_{a,b}^{\left(n\right)}\left(z\right),
\end{eqnarray*}
 in accordance with (\ref{eq:PBC}). With this definition, one may
also rewrite the $q$-fold degenerate Laughlin's state on the torus,
which is $\vartheta$-form is 
\begin{eqnarray*}
\psi_{\frac{1}{q}} & = & \mathcal{N\left(\tau\right)}e^{\i\pi\tau N_{\phi}\sum_{i}y_{i}^{2}}\elliptic{0,0}{\frac{q\sum z_{i}}{L}}{q\tau}\\
 &  & \quad\times\prod_{i<j}\elliptic{\frac{1}{2},\frac{1}{2}}{\frac{z_{i}-z_{j}}{L}}{\tau}^{q}.
\end{eqnarray*}
 In Weierstrass form this is the more compact

\begin{eqnarray}
\psi_{\frac{1}{q}} & = & \mathcal{N\left(\tau\right)}\weierstrauss q{0,0}{\sum_{i}z_{i}}\prod_{i<j}\weierstrauss 1{\frac{1}{2},\frac{1}{2}}{z_{i}-z_{j}}^{q}.\label{eq:Laughlin_sigma}
\end{eqnarray}
 The normalization factor is here chosen to be $\mathcal{N\left(\tau\right)}=\frac{\left[\sqrt{\tau_{2}}\eta^{2}\left(\tau\right)\right]^{\frac{qN_{e}}{2}}}{\eta\left(\tau\right)\frac{qN_{e}\left(N_{e}-1\right)}{2}+1}$
as suggested in Ref.\,\onlinecite{Read2009a}. This normalization
ensures that the Laughlin state transforms under $\S$-transformations
as $\psi_{\frac{1}{q}}\to\left(\frac{\tau}{\left|\tau\right|}\right)^{\frac{N_{e}q}{2}}\psi_{\frac{1}{q}}$.
This normalization is the correct one (up to $\tau$-independent scale
factors) as along as the torus is large enough\cite{Fremling2016b}.

\section{Girvin-Jach projection in $\tau$-gauge\label{sec:GJ-in-tau} }

In their work in Ref.\,\onlinecite{Girvin1984b}, Girvin and Jach
introduced the classic rule for LLL projection, namely that $\bar{z}\to2\partial_{z}$.
What might not be obvious is that this is a gauge dependent rule,
and is only guaranteed to hold in symmetric gauge. In this section
we review the Girvin-Jach projection trick\cite{Girvin1984b} and
then extend it to $\tau$-gauge. We begin by reminding ourselves of
the argument goes in symmetric gauge, before we turn to the $\tau$-gauge.
The ladder operators in symmetric gauge are 

\begin{eqnarray}
a_{s} & = & \sqrt{2}\left(\frac{z}{4}+\partial_{\bar{z}}\right)\nonumber \\
a_{s}^{\dagger} & = & \sqrt{2}\left(\frac{\bar{z}}{4}-\partial_{z}\right).\label{eq:Ladder_S}
\end{eqnarray}
 where the $s$ denotes the symmetric gauge choice. The equation for
$a^{\dagger}$ can be rewritten as $\bar{z}=\frac{4a_{s}^{\dagger}}{\sqrt{2}}+4\partial_{z}$,
which allows us to write 
\begin{align*}
\bar{z}e^{-\frac{z\bar{z}}{4}}f\left(z\right) & =\left(\frac{4a_{s}^{\dagger}}{\sqrt{2}}+4\partial_{z}\right)e^{-\frac{z\bar{z}}{4}}f\left(z\right)\\
 & =\frac{4a_{s}^{\dagger}}{\sqrt{2}}e^{-\frac{z\bar{z}}{4}}f\left(z\right)+e^{-\frac{z\bar{z}}{4}}\left(4\partial_{z}-\bar{z}\right)f\left(z\right),
\end{align*}

or equivalently 
\[
\bar{z}e^{-\frac{z\bar{z}}{4}}f\left(z\right)=\sqrt{2}a_{s}^{\dagger}e^{-\frac{z\bar{z}}{4}}f\left(z\right)+2e^{-\frac{z\bar{z}}{4}}\partial_{z}f\left(z\right).
\]
Applying the LLL projection kills the $a^{\dagger}$ term and we have
$P_{LLL}\bar{z}e^{-\frac{z\bar{z}}{4}}f\left(z\right)=2e^{-\frac{z\bar{z}}{4}}\partial_{z}f\left(z\right)$
which amounts to the famous rule\cite{Girvin1984b} $\bar{z}\to2\partial_{z}$,
where it is understood that the derivative does not act on the exponential
\textbf{$e^{-\frac{z\bar{z}}{4}}$}. Here, since $\left[a_{s}^{\dagger},\bar{z}\right]=0$
the argument can also be iterated to higher powers of $\bar{z}$ as
$\bar{z}^{n}\to\left(2\partial_{z}\right)^{n}$. In $\tau$-gauge,
due to (\ref{eq:Ladder_tau}), the same equations reads $\frac{\bar{\tau}}{2}Ly=\frac{a_{\tau}^{\dagger}}{\sqrt{2}}+\partial_{z}$,
which becomes the equation

\begin{align*}
 & \frac{\bar{\tau}}{2}LyG_{\tau}f\left(z\right)=\frac{a_{\tau}^{\dagger}}{\sqrt{2}}G_{\tau}f\left(z\right)+G_{\tau}\left(\partial_{z}+\frac{L\tau y}{2}\right)f\left(z\right).
\end{align*}
after acting on $G_{\tau}=e^{\i\pi\tau N_{\phi}y^{2}}$. The above
equation may be rewritten as

\[
\tau_{2}LyG_{\tau}f\left(z\right)=\frac{a_{\tau}^{\dagger}\i}{\sqrt{2}}G_{\tau}f\left(z\right)+G_{\tau}\left(\i\partial_{z}\right)f\left(z\right).
\]
After projection (and the $a_{\tau}^{\dagger}$ term is killed) this
becomes $\tilde{y}=\tau_{2}Ly\to\i\partial_{z}$, with the understanding
that $\partial_{z}$ does not act on the Gaussian factor $e^{\i\pi\tau N_{\phi}y^{2}}$.
The rule for $\tilde{y}$ can however not be extended directly to
higher powers of $\tilde{y}$ since $\left[\tilde{y},a_{\tau}^{\dagger}\right]\neq0$.
Instead due to this noncommutativity the projection rule reads
\begin{equation}
\tilde{y}^{n}\to\frac{1}{\left(-2\i\right)^{n}}H_{n}\left(\partial_{z}\right)\label{eq:y_tilde_LLL}
\end{equation}
 where $H_{n}$ is a Hermite polynomial. A proof and an extended discussion
can be found in Appendix \ref{app:y-LLL-projeciton}. We wish to stress
that since the $\plll$ operator only involves $a$ and $a^{\dagger}$
operators, that act between LL:s, it trivially commutes with the operators
within any LLL. This has the important consequence that if a wave
function satisfies the boundary conditions before projection, it will
automatically do so also after projection.

\subsection{LLL projection as an operator}

Here we develop a formalism where we view the LLL projection as an
operator action on holomorphic LLL wave functions. To be concrete,
we consider a general state (\emph{e.g. }basis state) $\phi_{n}^{\left(M\right)}$
in the $n$:th LL defined for $M$ fluxes, that is multiplied with
an arbitrary LLL wave function $\psi^{\left(N_{\phi}-M\right)}$ defined
for $N_{\phi}-M$ fluxes. The power of the $\tau$-gauge formalism
and reduced coordinates is that the product of $\phi_{n}^{\left(M\right)}\psi^{\left(N_{\phi}-M\right)}$
(when expressed in reduced coordinates) is automatically a proper
wave function at $N_{\phi}$ fluxes, since the different magnetic
lengths $\ell$ of the two wave functions are automatically renormalized.

The product $\phi_{n}^{\left(M\right)}\psi^{\left(N_{\phi}-M\right)}$
can now be written as 
\[
\phi_{n}^{\left(M\right)}\psi^{\left(N_{\phi}-M\right)}=e^{\i\pi\tau N_{\phi}y^{2}}f_{n}^{\left(M\right)}f^{\left(N_{\phi}-M\right)}
\]
 where $\psi^{\left(N_{\phi}-M\right)}=e^{\i\pi\tau\left(N_{\phi}-M\right)y^{2}}f^{\left(N_{\phi}-M\right)}$
is separated into its Gaussian and holomorphic factor, and the same
for $\phi_{n}^{\left(M\right)}=e^{\i\pi\tau My^{2}}f_{n}^{\left(M\right)}$.
When applying $\plll$ on this combined wave function we can use the
fact that only $\phi_{n}^{\left(M\right)}$ is non-holomorphic and
promote $f_{n}^{\left(M\right)}$ to a differential operator acting
on $f^{\left(N_{\phi}-M\right)}$ as 
\[
\plll\phi_{n}^{\left(M\right)}\psi^{\left(N_{\phi}-M\right)}=e^{\i\pi\tau N_{\phi}y^{2}}\hat{f}_{n}^{\left(M\right)}f^{\left(N_{\phi}-M\right)}.
\]
 The operator $\hat{f}_{n}^{\left(M\right)}$ can after some transformations
(see Appendix \ref{app:The-projection-operator}) be rewritten as 

\begin{eqnarray}
\hat{f}_{n}^{\left(M\right)} & = & \sum_{k=0}^{n}{n \choose k}\left(M\partial_{z}\right)^{n-k}\left[\left(-N_{\phi}\partial_{z}\right)^{k}f_{0}\right],\label{eq:f_form}
\end{eqnarray}
 where $f_{0}=e^{-\i\pi My^{2}}\frac{a^{n}}{\sqrt{n!}}\phi_{n}$ is
the LLL version of $f_{n}$, and where an scale factor of $\left(2\i\right)^{n}\frac{\mathcal{N}_{n}}{\mathcal{N}_{0}M^{n}}$
has been suppressed\footnote{This suppressed factor is irrelevant when one only considers products
states of CFs (or more generally when the CF superposition never mix
$\Lambda$-levels), but must be taken into account it $\Lambda$-level
mixing is needed.}. The derivative within square brackets acts only on $f_{0}$. 

We may symbolically write the operator in (\ref{eq:f_form}) as 

\begin{equation}
\hat{f}_{n}^{\left(M\right)}=\left(M\partial_{z}-N_{\phi}\tilde{\partial}_{z}\right)^{n}f_{0},\label{eq:f_title_n_form}
\end{equation}
 where the operator $\tilde{\partial}_{z}$ is understood to act only
on $f_{0}$ and thus has the property $\tilde{\partial}_{z}f_{0}f=f\tilde{\partial}_{z}f_{0}$.
We may also introduce the derivative operator $\hat{\partial}_{z}$
which does not act on $f_{0}$ at all and can be defined as $\hat{\partial}_{z}f_{0}f=f_{0}\hat{\partial}_{z}f$.
Using that these two operators have the identity $\partial_{z}=\tilde{\partial}_{z}+\hat{\partial}_{z}$
and that the three operators $\partial_{z}$, $\tilde{\partial}_{z}$,
$\hat{\partial}_{z}$ all commute, we may rewrite (\ref{eq:f_title_n_form})
as

\begin{equation}
\hat{f}_{n}=\left(N_{\phi}\hat{\partial}_{z}-\left(N_{\phi}-M\right)\partial_{z}\right)^{n}f_{0},\label{eq:f_hat_n_form}
\end{equation}
 and 
\begin{equation}
\hat{f}_{n}=\left(M\hat{\partial}_{z}-\left(N_{\phi}-M\right)\tilde{\partial}_{z}\right)^{n}f_{0},\label{eq:f_hat_title_n_form}
\end{equation}
 where especially (\ref{eq:f_hat_title_n_form}) will be useful later.
This is also the form that was found by PWJ. For brevity we will also
introduce the operator $\hat{D}=M\hat{\partial}_{z}-\left(N_{\phi}-M\right)\tilde{\partial}_{z}$
such that (\ref{eq:f_hat_title_n_form}) can be written in shorthand
as $\hat{f}_{n}=\hat{D}^{n}f_{0}$. 

\subsection{Periodic boundary conditions of $\hat{f}_{n}$}

To set the stage for the discussion of the PWJ projection in the later
sections we now prove that $\hat{f}_{n}$ indeed provides for periodic
boundary conditions. We know that an $A$ flux wave function $\psi^{\left(A\right)}=e^{\i\pi\tau Ay^{2}}f^{\left(A\right)}$
should obey the relation $t\left(\tau L\right)\psi^{\left(A\right)}=\psi^{\left(A\right)}$
(assuming p.b.c). Removing the factor $e^{\i\pi\tau Ay^{2}}$ and
the gauge factor $e^{\i2\pi Ax}$ we see that this implies that $\tilde{t}\left(\tau L\right)f^{\left(A\right)}=e^{-\i2\pi A\left(z+\frac{\tau}{2}\right)}f^{\left(A\right)}\tilde{t}\left(\tau L\right)$.
This means that the relation 
\begin{equation}
\left(\tilde{t}\left(\tau L\right)\hat{f}_{n}\right)e^{-\i2\pi\left(N_{\phi}-M\right)\left(z+\frac{\tau}{2}\right)}=e^{-\i2\pi N_{\phi}\left(z+\frac{\tau}{2}\right)}\hat{f}_{n}\label{eq:f_hat_pbc}
\end{equation}
 should hold for $\hat{f}_{n}$. Note here that $\left[\tilde{t}\left(\tau L\right),\hat{D}\right]=0$,
but that $\tilde{t}\left(\tau L\right)f_{0}=e^{-\i2\pi M\left(z+\frac{\tau}{2}\right)}f_{0}\tilde{t}\left(\tau L\right)$.
This means that when $\tilde{t}\left(\tau L\right)$ acts on $f_{0}$
is will produce the factor $e^{-\i2\pi M\left(z+\frac{\tau}{2}\right)}$.
This factor will the be acted upon by $\tilde{\partial}_{z}$ in eqn.
(\ref{eq:f_hat_title_n_form}), effectively causing the shift $\tilde{\partial}_{z}\to\tilde{\partial}_{z}-\i2\pi M$.
Likewise when $\tilde{t}\left(\tau L\right)$ acts on $f^{\left(N_{\phi}-M\right)}$
it produces the factor $e^{-\i2\pi\left(N_{\phi}-M\right)\left(z+\frac{\tau}{2}\right)}$,
which when pulled through $\hat{\partial}_{z}$ causes the shift $\hat{\partial}_{z}\to\hat{\partial}_{z}-\i2\pi\left(N_{\phi}-M\right)$.
Since the two shifts are simple constant they commute and we have
\begin{align}
 & \hat{D}=M\hat{\partial}_{z}-\left(N_{\phi}-M\right)\tilde{\partial}_{z}\nonumber \\
 & \to M\left(\hat{\partial}_{z}-\i2\pi\left(N_{\phi}-M\right)\right)\nonumber \\
 & -\left(N_{\phi}-M\right)\left(\tilde{\partial}_{z}-\i2\pi M\right)=\hat{D},\label{eq:D_to_D}
\end{align}
 when the exponentials are pulled through $\hat{D}$. This shows that
$\hat{D}$ is invariant and proves (\ref{eq:f_hat_pbc}).

We mention in passing that we may define $\hat{\phi}_{n}^{\left(M\right)}=e^{\i\pi\tau N_{\phi}y^{2}}\hat{f}_{n}^{\left(M\right)}e^{-\i\pi\tau\left(N_{\phi}-M\right)y^{2}}$
which is an operator that has proper operator boundary conditions.
This operator may be expressed as 
\begin{equation}
\hat{\phi}_{n}=\sum_{k=0}^{\left\lceil \frac{n}{2}\right\rceil }T\left(n,k\right)\chi^{k}D^{n-2k}\phi_{0}\label{eq:app_phi_expansion-1}
\end{equation}
 where $T\left(k,n\right)$ is the triangle of Bessel numbers\cite{OEIS_Bessel}
and $\chi=\frac{M\left(N-M\right)\tau}{4\i\tau_{2}}$. This has been
confirmed by Mathematica up to $n=8$, and we assume it holds for
general $n$. See Appendix (\ref{sec:pbc-Operators}) for details.

\section{Composite fermions on the Torus\label{sec:CFs_on_torus}}

In this section we briefly introduce the CF construction on the torus
at filling fraction $\nu=\frac{n}{2np+1}$ and discuss how the expected
degeneracy of $q=2pn+1$ comes about. A generic CF wave functions
may be written on the form 
\begin{eqnarray}
\psi_{CF} & = & \plll\chi_{n}\psi_{\nu=\frac{1}{2p}}\label{eq:CF_state}
\end{eqnarray}
 where $\chi_{n}$ is a Slater-determinant of occupied CF-orbitals
given by (\ref{eq:nLL-wfn}), where the CF-flux is $M$. If $\psi_{CF}$
represents a ground state at filing fraction $\nu=\frac{n}{2pn+1}$,
then $nM=N_{e}$ and $N_{\phi}=M+2pN_{e}$, meaning that the $n$
lowest CF $\Lambda$-levels are filled. As $\psi_{\nu=\frac{1}{2p}}$
contains a center of mass piece and a Jastrow factor (see eqn. (\ref{eq:Laughlin_sigma}))
we may pull the Jastrow factor into the determinant and write

\begin{eqnarray*}
\psi_{CF} & = & \plll\weierstrauss{2p}aZ\cdot\mathcal{A}\left\{ \prod_{j}\phi_{j}\left(z_{j}\right)\cdot J_{j}^{p}\left(z\right)\right\} .
\end{eqnarray*}

Here, $\mathcal{A}$ is an antisymmetrizer of the coordinates that
plays the same role as the determinant, and $J_{j}\left(z\right)=\prod_{k\ne j}\weierstrauss 1{\frac{1}{2},\frac{1}{2}}{z_{jk}}$.
The subscript $a$ on $\weierstrauss{2p}aZ$ is labeling one of the
$2p$ states of $\psi_{\nu=\frac{1}{2p}}$ and the subscript $j$
on $\phi_{j}$ contains for brevity both the LL-index and the orbital
index. We will later see that it is crucial for the PWJ projection
recipe that the CF state is a proper state. A proper CF state has
the property that there are no holes in the filling of the $\Lambda$-levels,
in the sense that if the orbital $\phi_{j,n}$ is occupied (with $n>0$),
then also the orbital $\phi_{j,n-1}$ is occupied.

\subsection{Notes on the multiplicity of the wave functions}

Here we mention for completeness how the correct degeneracy of the
CF states is counted. It is well known that for a LL with partial
filling $\nu=\frac{p}{q}$ ($p,q$ being relatively prime) every state
is at least $q$-fold degenerate on the torus\cite{Haldane1985a}
(with higher degeneracy for non-abelian states). To show this degeneracy
explicitly for the CF states, we make use of the many-body translation
operator commutations relations $T^{\left(A\right)}\left(\alpha\tau L\right)t^{\left(A\right)}\left(L\right)=e^{\i2\pi A\alpha}$,
where the $\left(A\right)$ denotes that the wave functions act on
$A$-flux wave functions. The many body operators are 
\[
T^{\left(A\right)}\left(\gamma\right)=\prod_{j=1}^{N_{e}}t_{j}^{\left(A\right)}\left(\gamma\right),
\]
where $t_{j}^{\left(A\right)}\left(\gamma\right)$ is the magnetic
translation operator in (\ref{eq:magnetic_translation}) acting on
coordinate $j$. We next define the translated state $\psi_{CF}^{\left(\alpha\right)}=T^{\left(N_{\phi}\right)}\left(\alpha\tau L\right)\psi_{CF}$.
If we assume that $\psi_{CF}$ has periodic boundary conditions then
$\psi_{CF}^{\left(\alpha\right)}$ will also have periodic boundary
conditions when $e^{\i2\pi N_{\phi}\alpha}=1$, which happens first
when $\alpha=\frac{1}{N_{\phi}}=\frac{n}{N_{e}\left(2pn+1\right)}$.
Naively one might expect that there should be $\frac{1}{\alpha}=N_{\phi}$
degenerate states from this argument, which is clearly wrong. To get
the correct counting, one has to also take into account that the trivial
cycle (\emph{i.e.} the cycle that sends $\psi_{CF}^{\left(\alpha\right)}=\psi_{CF}^{\left(0\right)}$)
is not $\alpha=1$, but is determined by the trivial cycles of $\psi_{\nu=1}^{2p}$
and $\chi_{n}$. The trivial cycle for $\psi_{\nu=1}$ is $\alpha^{\prime}=\frac{1}{N_{e}}$
since that cycles the states $\phi_{j,0}^{\left(N_{e}\right)}\to\phi_{j+1,0}^{\left(N_{e}\right)}$
leaving the $\psi_{\nu=1}$ invariant. In a similar manner, the trivial
cycle for $\chi_{n}$ is $\alpha^{\prime\prime}=\frac{1}{M}=\frac{n}{N_{e}}$
since it sends $\phi_{j,k}^{\left(M\right)}\to\phi_{j+1,k}^{\left(M\right)}$
in the determinant. We thus see that $\left(2pn+1\right)\alpha=n\alpha^{\prime}=\alpha^{\prime\prime}$
which shows that $2pn+1=q$ applications of $\alpha$ are needed to
obtain trivial cycles for the two sub-factors. This shows that the
degeneracy of $\psi_{CF}$ is $q=2pn+1$ as expected.

\section{Modified JK projection\label{sec:Modified-JK-projection} }

We now discuss the modification to (\ref{eq:f_hat_title_n_form})
that is necessary to obtain JK projected wave functions that respect
the periodic boundary conditions. In a naive implementation of the
JK projection we would move the projector into the determinant and
perform the LLL projection on each term of the determinant. On the
plane and sphere this is an uncomplicated procedure, but of the torus
this is highly nontrivial since the boundary conditions of the factor
$J_{j}\left(z\right)$ depends on the other $k\neq j$ coordinates.
Nevertheless we may be bold and stipulate that we can still use (\ref{eq:f_hat_n_form}),
and then hope for the best. In that case we first extract the Gaussian
factors and write 
\begin{eqnarray}
\psi_{JK} & = & \weierstrauss{2p}aZe^{\i\pi\tau\left(N_{\phi}\sum_{j}y_{j}^{2}-2pY^{2}\right)}\nonumber \\
 &  & \qquad\times\mathcal{A}\left\{ \prod_{j}\hat{f}_{j}\cdot F_{j}^{p}\left(z\right)\right\} ,\label{eq:Psi_JK}
\end{eqnarray}
 where now $\hat{f}_{j}$ only acts on the function 
\begin{equation}
F_{j}^{p}\left(z\right)=J_{j}^{p}e^{-\i\pi\tau p\sum_{k\neq j}\left(y_{j}-y_{k}\right)^{2}}.\label{eq:F_functoin}
\end{equation}

Here, and below, we use the abbreviations $Y=\sum_{j}y_{j}$, $X=\sum_{j}x_{j}$
and $Y_{j}=Y-y_{j}=\sum_{k\neq j}y_{k}$, $X_{j}=X-x_{j}=\sum_{k\neq j}x_{k}$.
Here the number of fluxes in $\hat{f}_{j}\cdot F_{j}^{p}\left(z\right)$
is $N_{\phi}=M+p\left(N_{e}-1\right)$ instead of $N_{\phi}=M+2pN_{e}$.
To see why this does not work, and also determine what does, we follow
the reasoning of PWJ and introduce a modification of $\hat{f}_{j}$
that is $\tilde{f}_{n}=\tilde{D}^{n}f_{0}$, where 
\begin{equation}
\tilde{D}=\alpha M\hat{\partial}_{z}-\left(N_{\phi}-M\right)\tilde{\partial}_{z}=\alpha M\hat{\partial}_{z}-2pN_{e}\tilde{\partial}_{z}.\label{eq:D_tilde_operator}
\end{equation}
For $\alpha=1$ then $\tilde{D}=\hat{D}$ and $\hat{f}_{j}=\tilde{f}_{j}$,
but we will soon see that the choice $\alpha=2$ will be necessary.
We begin with reviewing the relevant transformations. Acting with
$\tilde{t}_{j}\left(\tau L\right)$ on $F_{l}$ produces (after we
have dropped some constant factors)
\begin{align}
\tilde{t}_{j}\left(\tau L\right)F_{j}\left(z\right) & \propto e^{-\i2\pi\prod_{k\ne j}\left(z_{j}-z_{k}\right)}\prod_{k\neq j}\elliptic 1{z_{j}-z_{k}}{\tau}\tilde{t}_{j}\left(\tau L\right)\nonumber \\
 & =e^{-\i2\pi\left[\left(N_{e}-1\right)z_{j}-Z_{j}\right]}F_{j}\tilde{t}_{j}\left(\tau L\right)\label{eq:tj_on_Fj}
\end{align}
 and 
\begin{align}
\tilde{t}_{j}\left(\tau L\right)F_{l\neq j}\left(z\right) & \propto e^{-\i2\pi\left(z_{l}-z_{j}\right)}\prod_{k\neq l}\elliptic 1{z_{l}-z_{k}}{\tau}\tilde{t}_{j}\left(\tau L\right)\label{eq:tj_on_Fl}\\
 & =e^{-\i2\pi\left(z_{l}-z_{j}\right)}F_{l\neq j}\tilde{t}_{j}\left(\tau L\right).\nonumber 
\end{align}
 depending on if $j=l$ or not. We now apply the translation operator
$\tilde{t}_{j}\left(\tau L\right)$ on $\tilde{f}_{j}F_{j}$. For
brevity we suppress the factors of $e^{-\i2\pi\left[\left(N_{e}-1\right)z_{j}-Z_{j}\right]}$
and $e^{-\i2\pi\left(z_{l}-z_{j}\right)}$ coming from (\ref{eq:tj_on_Fj})
and (\ref{eq:tj_on_Fl}) as well as the phase $e^{\i2\pi z_{j}M}$
coming from $f_{0}^{\left(M\right)}$. We obtain, by an analogs calculation
to the one in (\ref{eq:D_to_D}) that\begin{widetext} 

\begin{align*}
\hat{f}_{n}^{\left(M\right)}F_{j}^{p} & \to\left(-2N_{e}p\left(\tilde{\partial}_{z_{j}}-\i2\pi M\right)+\alpha M\left(\hat{\partial}_{z_{j}}-\i2\pi p\left(N_{e}-1\right)\right)\right)^{n}f_{0}F_{j}^{p}\\
 & =\left(\tilde{D}+\i2\pi pM\left[2N_{e}-\alpha\left(N_{e}-1\right)\right]\right)^{n}f_{0}F_{j}^{p}.
\end{align*}
 \end{widetext} for $j=l$. For $j\neq l$ we instead have

\begin{align*}
\hat{f}_{n}^{\left(M\right)}F_{l\neq j} & \to\left(-2N_{e}p\tilde{\partial}_{z_{j}}+\alpha M\left(\hat{\partial}_{z_{j}}+\i2\pi p\right)\right)^{n}f_{0}F_{l}^{p}\\
 & =\left(\tilde{D}+\alpha pM\i2\pi\right)^{n}f_{0}F_{l}^{p}.
\end{align*}
The important observation here is that both $\alpha pM\i2\pi$ and
$\i2\pi pM\left[2N_{e}-\alpha\left(N_{e}-1\right)\right]$ are constants,
but they are only equal when $\alpha=2$. It is crucial that the transformation
$\tilde{D}^{n}\to\left(\tilde{D}+\text{const}\right)^{n}$ is the
same for all coordinates, since the shift $\i\pi4pM$ can then be
removed by row addition if the CF state is a proper state. Otherwise
the cancellation will not work.

As a minimal example lets consider the simple case of a determinant
consisting of only $N_{e}=2$ particles; one in the $n=0$ LL and
one in the $n=1$ LL. The entries in (\ref{eq:Psi_JK}) are then $DfF$
and $fF$, which gives determinant 
\[
\left|\begin{array}{cc}
D_{1}f_{1}F_{1} & D_{2}f_{2}F_{2}\\
f_{1}F_{1} & f_{2}F_{2}
\end{array}\right|,
\]
 where the subscripts labels the coordinates of the two particles.
If we assume that $D_{1}\to D_{1}+\alpha$ and $D_{2}\to D_{2}+\beta$
under the action of $t_{1}\left(\tau L\right)$, we then have 

\begin{align*}
\left|\begin{array}{cc}
D_{1}f_{1}F_{1} & D_{2}f_{2}F_{2}\\
f_{1}F_{1} & f_{2}F_{2}
\end{array}\right| & \to\left|\begin{array}{cc}
\left(D_{1}+\alpha\right)f_{1}F_{1} & \left(D_{2}+\beta\right)f_{2}F_{2}\\
f_{1}F_{1} & f_{2}F_{2}
\end{array}\right|\\
 & =\left|\begin{array}{cc}
D_{1}f_{1}F_{1} & \left(D_{2}+\left(\beta-\alpha\right)\right)f_{2}F_{2}\\
f_{1}F_{1} & f_{2}F_{2}
\end{array}\right|.
\end{align*}
 It is evident that the determinant is only invariant under the transformation
if $\alpha=\beta$.

\section{Modular Covariance\label{sec:Modular-Covariance}}

We are now in a positions to study the modular covariance properties
of the PWJ wave functions. For this purpose (and to simplify the discussion
somewhat) we assume that we are considering one of the CF ground states
at filling fraction $\nu=\frac{n}{2pn+1}$. That is, we assume that
we have a state with $n$ filled $\Lambda$-levels, and everything
above unoccupied. In this work we will focus on the $\S$-transform,
$\tau\to-\frac{1}{\tau}$ since that is the more complicated of the
two.

Before we deal with the many-body state, let us review how single
particle orbitals transform under the $\S$-transform. An $S$-transform
sends $\tau\to-\frac{1}{\tau}$ and affects the LLL single particle
orbitals from (\ref{eq:LLL-wfn}) as\begin{widetext} 

\begin{equation}
\phi_{i,0}^{\left(M\right)}\left(x,y,-\frac{1}{\tau}\right)=e^{-\i2\pi Myx}\sqrt{\frac{\tau}{\left|\tau\right|}}\frac{1}{\sqrt{M}}\sum_{k}e^{\i2\pi\frac{ki}{M}}\phi_{k,0}^{\left(M\right)}\left(-y,x,\tau\right).\label{eq:LLL-S-transform}
\end{equation}
 The higher order higher LL orbitals in (\ref{eq:nLL-wfn}) similarly
transform as 

\begin{equation}
\phi_{i,n}^{\left(M\right)}\left(x,y,-\frac{1}{\tau}\right)=e^{-\i2\pi Myx}\left(\frac{\tau}{\left|\tau\right|}\right)^{n+\frac{1}{2}}\frac{1}{\sqrt{M}}\sum_{k}e^{\i2\pi\frac{ki}{M}}\phi_{k,n}^{\left(M\right)}\left(-y,x,\tau\right).\label{eq:nLL-S-transform}
\end{equation}
 \end{widetext} In the above equations we note that $\tau\to-\frac{1}{\tau}$
effectively sends $y\to x\to-y$ and maps $\phi_{i,n}^{\left(M\right)}$
into a Fourier sum $\sum_{k}e^{\i2\pi\frac{ki}{M}}\phi_{k,n}^{\left(M\right)}$.
We may also identify the factor $e^{-\i2\pi Myx}$ as the gauge transformation
related with the coordinate change and $\left(\frac{\tau}{\left|\tau\right|}\right)^{n+\frac{1}{2}}$
can be interpreted as the conformal weight of the orbital. The extra
factors of $\frac{\tau}{\left|\tau\right|}$ for $n>0$ can be understood
by noting that the derivative operator $\partial_{z}^{\left(x,y,\tau\right)}$
transforms as 

\begin{eqnarray*}
\partial_{z}^{\left(x,y,-\frac{1}{\tau}\right)} & = & \frac{\left|\tau\right|}{\bar{\tau}L2\i\tau_{2}}\left(\bar{\tau}\partial_{y}+\partial_{x}\right)\\
 & = & \left(\frac{\tau}{\left|\tau\right|}\right)\partial_{z}^{\left(-y,x,\tau\right)}.
\end{eqnarray*}
 In this calculation we used that $\tau_{2}\to\frac{\tau_{2}}{\left|\tau\right|^{2}}$,
$L=\sqrt{\frac{2\pi N_{\phi}}{\tau_{2}}}\to\left|\tau\right|\sqrt{\frac{2\pi N_{\phi}}{\tau_{2}}}=\left|\tau\right|L$
and that $\bar{\tau}\to-\frac{1}{\bar{\tau}}=\frac{-\tau}{\left|\tau\right|^{2}}$.
Thus, the  ladder operator $a_{\tau}^{\dagger}\left(x,y,\tau\right)=-\sqrt{2}\left(\partial_{z}-\frac{\bar{\tau}}{2}Ly\right)$
transforms as 

\begin{eqnarray}
 &  & a_{\tau}^{\dagger}\left(x,y,-\tau\right)=-\sqrt{2}\left(\frac{\tau}{\left|\tau\right|}\right)\left(\partial_{z}^{\left(-y,x,\tau\right)}+\frac{1}{2}Ly\right)\nonumber \\
 &  & \qquad=e^{-\i2\pi Myx}\left(\frac{\tau}{\left|\tau\right|}\right)a_{\tau}^{\dagger}\left(-y,x,\tau\right)e^{\i2\pi Myx}.\label{eq:ladder_transform}
\end{eqnarray}
 By applying (\ref{eq:ladder_transform}) and (\ref{eq:LLL-S-transform})
to $\phi_{j,n}=a_{\tau}^{\dagger n}\phi_{j,0}$ then (\ref{eq:nLL-S-transform})
is directly obtained. From this also follows that $J_{j}^{p}$, which
is used in the CF construction will transform trivially under $\tau\to-\frac{1}{\tau}$
since they contain a product of $\weierstrauss 1{\frac{1}{2},\frac{1}{2}}{z_{ij}}$
which are an $M=1$ representation.

\subsection{A modular invariant CF wave function\label{subsec:A-modular-invariant}}

To make the discussion in the following subsection a little bit cleaner,
we spend some time in this section defining a CF wave function that
transforms trivially under $\S$-transforms in its unprojected form.
We do this, since if we can find one wave function $\psi$ that transforms
trivially under $S$ we can then build the family of $q$-fold degenerate
wave functions from this template, as eigenstates of either $T_{1}=T\left(\frac{L}{N_{\phi}}\right)$
or $T_{2}=T\left(\frac{\tau L}{N_{\phi}}\right)$.

In practice, if we assume that $\psi$ is a wave function that is
modular invariant, we may define the states $\psi_{j}=T_{2}^{j}P_{1,0}\psi$
and $\varphi_{l}=T_{1}^{-l}P_{2,0}\psi$. Here 
\[
P_{m,l}=\frac{1}{q}\sum_{j}e^{\i2\pi\frac{-lj}{q}}T_{m}^{j},
\]
 is a projector onto the basis defined by $T_{m}$ and it satisfies
$\sum_{l=1}^{q}P_{m,l}=1$ and $P_{m,l}P_{m,k}=P_{m,l}\delta_{l,k}$.
Since $T_{1}T_{2}=T_{2}T_{1}e^{\i2\pi\frac{p}{q}}$ we have that 
\begin{align*}
\psi_{l} & \propto\frac{1}{\sqrt{q}}\sum_{j}e^{\i2\pi\frac{-plj}{q}}\varphi_{j}\\
\varphi_{j} & \propto\frac{1}{\sqrt{q}}\sum_{j}e^{\i2\pi\frac{plj}{q}}\psi_{j}
\end{align*}
 where the $\propto$ is inserted since $\psi_{l}$ and $\varphi_{j}$
might not be properly normalized with respect to each other. Now,
by applying the $S$-transform, which transforms $T_{1}\to T_{2}\to T_{2}^{-1}$,
we find that 
\begin{align*}
\psi_{l} & =T_{2}^{j}P_{1,0}\psi\\
 & \to T_{1}^{-j}P_{2,0}\psi\\
 & =\varphi_{l}\propto\frac{1}{\sqrt{q}}\sum_{j}e^{\i2\pi\frac{plj}{q}}\psi_{j}
\end{align*}
 which shows that the set of $q$ wave functions $\psi_{l}$ is closed
under $\S$. It thus remains to be seen that $\psi$ transforms trivially
under $\S$.

\subsection{Unprojected CF\label{subsec:Unprojected-CF}}

According to the argument of the previous section, it is sufficient
to show modular covariance if we can find one CF-wave function that
is invariant under the $\S$-transform. For this purpose we note that
if we choose $\psi_{\nu=1}^{2p}$ instead of $\psi_{\nu=\frac{1}{2p}}$
in (\ref{eq:CF_state}) then the center of mass part and the Jastrow
factors $\sigma_{\frac{1}{2},\frac{1}{2}}\left(z_{ij}\right)$ are
all manifestly invariant under these transformations (up to constant
factors and phases). The determinant $\chi_{n}$ can be made invariant
in two different but equivalent ways. The first is is argue that if
one fills a $\Lambda$-levels completely, it will also be filled after
the $S$-transform, thus ensuring the invariance. The second, which
will make the later discussion of the PWJ projection much cleaner,
is to build $\chi_{n}$ from orbitals that themselves are invariant
under the $\S$-transform. 

By choosing the $\chi_{n}$ orbitals from the Lattice coherent states
\begin{equation}
\rho_{n,m}\left(z\right)=\sigma_{\frac{1}{2},\frac{1}{2}}\left(z-\frac{1}{M}\left(n+\tau m\right)\right)^{M},\label{eq:Lattice_states}
\end{equation}
 one can ensure that each orbital is invariant under $\S$. These
states where introduced by Haldane in Ref.\,\onlinecite{Haldane1985b}
as a possible way to construct maximally localized wave functions
and where later studied in detail in Ref.\,\onlinecite{Fremling2013}.
They have the property that they have all their zeroes at the same
position $z=L\left(n+\tau m\right)$, and transform as $\rho_{n,m}\to\rho_{m,-n}$
under modular transformations. By constructing the states $\tilde{\rho}_{n,m}=\rho_{n,m}+\rho_{m,-n}+\rho_{-n,-m}+\rho_{-m,n}$
we ensure that all the orbitals transform trivially under the $\S$-transform.
These are examples of eigenstates for certain finite rotations. The
LCS forms an over complete $M\times M$ lattice of states and there
are thus roughly $M^{2}/4$ acceptable choices for $\tilde{\rho}_{n,m}$.
Since $M^{2}/4>M$ these states are enough to fill the lowest of the
$\Lambda$-levels and thus all also of the higher $\Lambda$-levels
by the action of raising operators.

\subsection{Exactly projected CF}

To prove that the exactly projected states have good modular properties,
it is sufficient to show that the modular transformation commutes
with the projector $\plll$. This is straight forward due to equations
(\ref{eq:LLL-S-transform}) and (\ref{eq:nLL-S-transform}). These
equations namely show that the modular transformation never mixes
states between landau levels, and thus trivially commutes with the
Landau level projection. 

A more formal proof of the same is to note that $\plll$ can formally
we written as $\plll=\prod_{n=1}^{\infty}\left(1-\frac{a^{\dagger}a}{n}\right)$
where $a^{\dagger},$$a$ are the operators in (\ref{eq:Ladder_tau}).
Using the result from (\ref{eq:ladder_transform}) we see that $a_{\tau}^{\dagger}a_{\tau}\to e^{-\i2\pi Myx}a_{\tau}^{\dagger}a_{\tau}e^{\i2\pi Myx}$
only contains the overall gauge transformation $e^{\i2\pi Myx}$,
and so $\plll\to e^{-\i2\pi Myx}\plll e^{\i2\pi Myx}$ under the $\S$
transform. This shows that $\plll$ commutes with $\S$ up to the
ever present gauge transformation.

\subsection{PWJ projected CF}

We now turn our attention to the PWJ projected CF state, where we
are especially interested in the transformation properties of $\tilde{f}_{n}$
and $F_{j}^{p}$ as defined in (\ref{eq:F_functoin}) and (\ref{eq:D_tilde_operator}).
We here assume, following the discussion in (\ref{subsec:A-modular-invariant})
and (\ref{subsec:Unprojected-CF}) that $f_{0}$ is chosen from the
set of Lattice coherent states (\ref{eq:Lattice_states}). For $f_{j}\equiv f_{0}\left(z_{j}\right)$
and $F_{j}$ we have the respective transformations (again with constant
faces removed) 
\[
f_{j}^{\left(M\right)}\to e^{\i\pi\tau Mz_{j}^{2}}f_{j}^{\left(M\right)}
\]
 and 
\begin{align*}
F_{j} & \to e^{\i\tau\pi\sum_{l}\left(z_{j}-z_{l}\right)^{2}}F_{j}=e^{\i\tau\pi\left(N_{e}z_{j}^{2}-2z_{j}Z+\sum_{l}z_{l}^{2}\right)}F_{j}\\
 & =e^{\i\tau\pi\left(\left(N_{e}-1\right)z_{j}^{2}-2z_{j}Z_{j}+\sum_{l\neq j}z_{l}^{2}\right)}F_{j}.
\end{align*}
The combined transformation is thus
\begin{align*}
 & \tilde{D}_{j}^{n}f_{j}F_{j}^{p}\to\tilde{D}_{j}^{n}\left(e^{\i\pi\tau Mz_{j}^{2}}f_{j}\right)\\
 & \qquad\times\left(e^{\i\tau\pi p\left(\left(N_{e}-1\right)z_{j}^{2}-2z_{j}Z_{j}+\sum_{l\neq j}z_{l}^{2}\right)}F_{j}^{p}\right).
\end{align*}
 Let us first consider the simplest case of $n=1$ where we define
$\gamma=e^{\i\pi\tau Mz_{j}^{2}}e^{\i\tau\pi p\left(\left(N_{e}-1\right)z_{j}^{2}-2z_{j}Z_{j}+\sum_{l\neq j}z_{l}^{2}\right)}$.
This yields

\begin{align*}
D_{j}f_{j}F_{j}^{p} & \to\gamma^{-1}\left(-2pN_{e}\tilde{\partial}_{z_{j}}+\boldsymbol{2}M\hat{\partial}_{z_{j}}\right)\left(e^{\i\pi\tau Mz_{j}^{2}}f_{j}\right)\\
 & \times\left(e^{\i\tau\pi p\left(\left(N_{e}-1\right)z_{j}^{2}-2z_{j}Z_{j}+\sum_{l\neq j}z_{l}^{2}\right)}F_{j}^{p}\right)\\
 & =-2pN_{e}\left(\tilde{\partial}_{z_{j}}+\i2\pi\tau Mz_{j}\right)f_{j}F_{j}^{p}\\
 & \qquad+\boldsymbol{2}M\left(\hat{\partial}_{z_{j}}+\i\tau\pi2p\left(\left(N_{e}-1\right)z_{j}-Z_{j}\right)\right)f_{j}F_{j}^{p}\\
 & =D_{j}f_{j}F_{j}^{p}-\i4\pi\tau MpZf_{j}F_{j}^{p},
\end{align*}
where we see an extra term $-\i\tau\pi4MpZf_{j}F_{j}^{p}$ appearing.
This term can then be removed under row addiction of the determinant.
This is since it is proportional to $Zf_{j}F_{j}^{p}$ and $Z$ is
independent of the $j$ index. 

For general $n$ we cannot use the trick employed above since $\left[\partial_{z},\left[\partial_{z},z^{2}\right]\right]=2\ne0$,
which means that the factors $\tilde{\partial}_{z_{j}}\to\tilde{\partial}_{z_{j}}+\i2\pi\tau Mz_{j}$
and $\hat{\partial}_{z_{j}}\to\hat{\partial}_{z_{j}}+\i\tau\pi2p\left(\left(N_{e}-1\right)z_{j}-Z_{j}\right)$
can only in the $n=1$ case be direly combined to $\tilde{D}\to\tilde{D}-\i\tau\pi4MpZ$.
For the $n=2$ case, one may after some algebra conclude that 

\begin{align*}
\tilde{D}^{2} & \to\tilde{D}^{2}\\
 & +\i\pi\tau8MpZ\tilde{D}\\
 & -16\left(\pi\tau Mp\right)^{2}Z^{2}\\
 & -8\i\pi\tau Mp\left(M\left(N_{e}-1\right)+N_{e}^{2}p\right).
\end{align*}
Here we see that we still only get terms that depend on $Z$ and $\tilde{D}$,
and they can all be removed by row addition. By Mathematica calculations
we have confirm up to $n=10$, and we belie it holds in general, that
the general transformation that takes place is 
\begin{align*}
D^{n} & \to\sum_{k=0}^{n}\sum_{l=0}^{\left\lfloor \frac{k}{2}\right\rfloor }A_{k,l}Z^{k-2l}D^{n-k}\alpha^{k}\beta^{l}.
\end{align*}
 The constants in the expression are $\alpha=\i\pi\tau4Mp$, $\beta=\frac{M\left(N_{e}-1\right)+N_{e}^{2}p}{\i\pi\tau2Mp}$
and $A_{k,l}$ is defined as 
\begin{align*}
A_{k,0} & ={n \choose k}\\
A_{k,l} & =-\frac{A_{2l,l-1}}{l}{n-2l \choose k-2l}.
\end{align*}
Again, since the extra terms that are generated are all proportional
to powers of $Z$ and $\tilde{D}$, they can all be removed by row-addition
in the determinant is the CF state is proper. This proves that the
PWJ wave functions transform nicely under modular transformations.

\section{Discussion\label{sec:Discussion}}

In this paper we have shown explicitly that the CF wave functions
have proper modular properties on the torus. As part of this work
we have also reformulated the PWJ method in $\tau$-tau gauge, which
is the natural gauge choice for the tours. We have along the way exposed
a series of analytical expressions for the projected states that we
hope will be useful for future studies of composite fermions on the
torus. One limitation of the original PWJ formulation is that it is
not applicable for reverse flux states, and we especially hope that
this is a step in extending the PWJ method to this class of CF wave
functions.

\begin{acknowledgments}

We thank J. K. Jain, S. Pu and H. Hansson for enlightening discussions
and for the encouragement to write this manuscript. We gratefully
acknowledge financial support from Science Foundation Ireland Principal
Investigator Award 12/IA/1697 and from Netherlands Organisation for
Scientific Research (NWO).

\end{acknowledgments}

\bibliographystyle{alpha}
\bibliography{References}

\appendix

\section{The LLL projection of $\tilde{y}$\label{app:y-LLL-projeciton}}

We know form the work of Grivin and Jach\cite{Girvin1984b} that we
may write the LLL projection as 
\begin{equation}
\plll\bar{z}G_{s}f_{s}\left(z\right)=G_{s}\left(2\partial_{z}\right)f_{s}\left(z\right),\label{eq:app_GJ_rule}
\end{equation}
 where $G_{s}=e^{-\frac{z\bar{z}}{4}}$ and $f_{s}\left(z\right)$
is a polynomial in $z$. Any wave function in symmetric gauge can
be transformed into $\tau$-gauge with the action of the unitary operator
$U_{s\to\tau}=e^{-\i\pi N_{\phi}xy}$ in such a way that 
\[
U_{s\to\tau}G_{s}f_{s}\left(z\right)=G_{\tau}f_{\tau}\left(z\right),
\]
 where $G_{\tau}=e^{\i\pi\tau N_{\phi}y^{2}}$ and $f_{\tau}\left(z\right)=e^{-\frac{z^{2}}{4}}f_{s}\left(z\right)$
is also a holomorphic polynomial. Technically, also $\plll$ is gauge
dependent but we suppress that in the analysis below. Applying $U_{s\to\tau}$
to the left and right hand sides of eqn. (\ref{eq:app_GJ_rule}) now
gives 

\begin{equation}
\plll\bar{z}G_{\tau}f_{\tau}\left(z\right)=G_{\tau}e^{-\frac{z^{2}}{4}}\left(2\partial_{z}\right)e^{\frac{z^{2}}{4}}f_{\tau}\left(z\right),\label{eq:app_GJ_rule-tau}
\end{equation}
 which after pulling the $\partial_{z}$ through the $e^{\frac{z^{2}}{4}}$
gives $\plll\bar{z}G_{\tau}f_{\tau}\left(z\right)=G_{\tau}\left(2\partial_{z}+z\right)f_{\tau}\left(z\right)$.
Finally after moving $G_{\tau}zf_{\tau}\left(z\right)$ to the left
hand side and using that $\plll G_{\tau}zf_{\tau}\left(z\right)=G_{\tau}zf_{\tau}\left(z\right)$
while identifying $\bar{z}-z=-2\i\tilde{y}$, we have 

\begin{equation}
\plll\tilde{y}G_{\tau}f_{\tau}\left(z\right)=G_{\tau}\left(\i\partial_{z}\right)f_{\tau}\left(z\right),\label{eq:app_GJ_rule-tau-1}
\end{equation}
 just as in the main text. The generalization to higher powers of
$\tilde{y}$ is straight forward since we can write\begin{widetext}
\begin{align}
\plll\tilde{y}^{n}G_{\tau}f_{\tau}\left(z\right) & =\frac{1}{\left(-2\i\right)^{n}}\plll\left(\bar{z}-z\right)^{n}G_{\tau}f_{\tau}\left(z\right)\nonumber \\
 & =\frac{1}{\left(-2\i\right)^{n}}\plll\sum_{k=0}^{n}{n \choose k}\bar{z}^{k}\left(-z\right)^{n-k}G_{\tau}f_{\tau}\left(z\right)\nonumber \\
 & =\frac{1}{\left(-2\i\right)^{n}}G_{\tau}\sum_{k=0}^{n}{n \choose k}e^{-\frac{z^{2}}{4}}\left(2\partial_{z}\right)^{k}e^{\frac{z^{2}}{4}}\left(-z\right)^{n-k}f_{\tau}\left(z\right)\nonumber \\
 & =\frac{1}{\left(-2\i\right)^{n}}G_{\tau}\sum_{k=0}^{n}{n \choose k}\left(2\partial_{z}+z\right)^{k}\left(-z\right)^{n-k}f_{\tau}\left(z\right).\label{eq:app_proj_y_exp}
\end{align}
 \end{widetext}For the second row we expanded $\left(\bar{z}-z\right)^{n}$,
for the third we used the rule $\bar{z}^{k}\to\left(2\partial_{z}\right)^{k}$
and for the last row that $e^{-\frac{z^{2}}{4}}\left(2\partial_{z}\right)^{k}e^{\frac{z^{2}}{4}}=\left(2\partial_{z}+z\right)^{k}$.
We will now prove that eqn. \ref{eq:app_proj_y_exp} can be rewritten
as the more elegant 
\[
\plll\tilde{y}^{n}G_{\tau}f_{\tau}\left(z\right)=\frac{1}{\left(-2\i\right)^{n}}G_{\tau}H_{n}\left(\partial_{z}\right)f_{\tau}\left(z\right)
\]
 where $H_{n}\left(x\right)$ is a Hermite polynomial. The proof uses
that the Hermite polynomial satisfies the relation $H_{n+1}\left(x\right)=2xH_{n}\left(x\right)-H_{n}^{\prime}\left(x\right)$.
Since $H_{n}$ has an operator $\partial_{z}$ as argument, we can
implement the derivative with respect to $\partial_{z}$ as $\frac{\partial}{\partial_{\partial_{z}}}H_{n}\left(\partial_{z}\right)=\left[H_{n}\left(\partial_{z}\right),z\right]$.
We then get the equation
\begin{equation}
H_{n+1}\left(\partial_{z}\right)=2\partial_{z}H_{n}\left(\partial_{z}\right)-\left[H_{n}\left(\partial_{z}\right),z\right]\label{eq:app_Hn_rec_rel}
\end{equation}
 where we propose that 
\begin{equation}
H_{n}\left(\partial_{z}\right)\overset{?}{=}\sum_{k=0}^{n}{n \choose k}\left(2\partial_{z}+z\right)^{k}\left(-z\right)^{n-k}\label{eq:app_H_n_exp}
\end{equation}
 is a solution. We construct a proof by induction. First we show that
$H_{1}\left(\partial_{z}\right)=\left(-z\right)+\left(2\partial_{z}+z\right)=2\partial_{z}$
is trivially true. After some algebra we can show that (\ref{eq:app_proj_y_exp})
satisfies the recursion relation (\ref{eq:app_Hn_rec_rel}). This
is since\begin{widetext} 
\begin{align*}
 & 2\partial_{z}H_{n}\left(\partial_{z}\right)-\left[H_{n}\left(\partial_{z}\right),z\right]\\
 & =\sum_{k=0}^{n}{n \choose k}\left[2\partial_{z}\left(2\partial_{z}+z\right)^{k}\left(-z\right)^{n-k}+\left(2\partial_{z}+z\right)^{k}\left(-z\right)^{n+1-k}+z\left(2\partial_{z}+z\right)^{k}\left(-z\right)^{n-k}\right]\\
 & =\sum_{k=0}^{n}{n \choose k}\left(2\partial_{z}+z\right)^{k+1}\left(-z\right)^{n-k}+\sum_{k=0}^{n}{n \choose k}\left(2\partial_{z}+z\right)^{k}\left(-z\right)^{n+1-k}\\
 & =\left\{ \sum_{k=1}^{n+1}{n \choose k-1}+\sum_{k=0}^{n}{n \choose k}\right\} \left(2\partial_{z}+z\right)^{k}\left(-z\right)^{n+1-k}\\
 & =\sum_{k=0}^{n+1}{n+1 \choose k}\left(2\partial_{z}+z\right)^{k}\left(-z\right)^{n+1-k}=H_{n+1}\left(\partial_{z}\right),
\end{align*}
 \end{widetext}where on line four we used that ${n \choose n+1}={n \choose -1}=0$.
This concludes the proof.

\section{The projection operator\label{app:The-projection-operator}}

In this section we investigate the effect of the GJ trick on the $n$:th
LL wave function $\phi_{j,n}^{\left(M\right)}$ as defined in (\ref{eq:nLL-wfn}),
where it is also understood that this is always multiplied with a
$N_{\phi}-M$ flux wave function. If we strip of the leading Gaussian
we have the wave function 

\begin{eqnarray}
f_{j,n}^{\left(M\right)} & = & \mathcal{N}_{n}\sum_{k\in\mathbb{Z}+\frac{j}{N_{\phi}}}e^{\i\pi M\tau k^{2}}H_{n}\left(\tilde{y}+\tau_{2}Lk\right)e^{\i2\pi Mk\frac{z}{L}}.\label{eq:app_f_n}
\end{eqnarray}
 As mentioned in the previous section, we cannot simply replace $\tilde{y}\to\i\partial_{z}$,
but the rule is rather that $\tilde{y}^{n}\to\frac{1}{\left(-2\i\right)^{n}}H_{n}\left(\partial_{z}\right)$.
By expanding the Hermite polynomial in powers of $\tilde{y}+\tau_{2}Lk$
we have 

\begin{align*}
 & H_{n}\left(\tilde{y}+\tau_{2}Lk\right)=\sum_{m=0}^{\left\lfloor \frac{n}{2}\right\rfloor }g_{n,m}\left(\tilde{y}+\tau_{2}Lk\right)^{n-2m}\\
 & \quad=\sum_{m=0}^{\left\lfloor \frac{n}{2}\right\rfloor }g_{n,m}\sum_{r=0}^{n-2m}{n-2m \choose r}\tilde{y}^{n-2m-r}\left(\tau_{2}Lk\right)^{r},
\end{align*}
 where we used the expansion $H_{n}\left(x\right)=\sum_{m=0}^{\left\lfloor \frac{n}{2}\right\rfloor }g_{n,m}x^{n-2m}$
and $g_{n,m}=\frac{n!\left(-1\right)^{m}2^{n-2m}}{m!\left(n-2m\right)!}$.
We note that we can write $\left(\tau_{2}Lk\right)^{r}e^{\i2\pi Mk\frac{z}{L}}\cdot1=\left(-\i\partial_{z}\frac{N_{\phi}}{M}\right)^{r}e^{\i2\pi Mk\frac{z}{L}}\cdot1$.
This allows us to write (\ref{eq:app_f_n}) as 

\begin{eqnarray*}
\hat{f}_{j,n}^{\left(M\right)} & = & \frac{\mathcal{N}_{n}}{\mathcal{N}_{0}}\sum_{m=0}^{\left\lfloor \frac{n}{2}\right\rfloor }g_{n,m}\sum_{r=0}^{n-2m}{n-2m \choose r}\times\\
 &  & \quad\times\tilde{y}^{n-2m-r}\left[\left(-\i\partial_{z}\frac{N_{\phi}}{M}\right)^{r}f_{j,0}\right],
\end{eqnarray*}
 where the $\left[\ldots\right]$ signifies that the derivative does
not act outside of the square bracket box. After the projection step
this becomes 

\begin{eqnarray*}
\hat{f}_{j,n}^{\left(M\right)} & = & \frac{\mathcal{N}_{n}}{\mathcal{N}_{0}}\sum_{m=0}^{\left\lfloor \frac{n}{2}\right\rfloor }g_{n,m}\sum_{r=0}^{n-2m}{n-2m \choose r}\times\\
 &  & \quad\times\frac{H_{n-2m-r}\left(\partial_{z}\right)}{\left(-2\i\right)^{n-2m-r}}\left[\left(-\partial_{z}\frac{N_{\phi}}{M}\right)^{r}f_{j,0}\right].
\end{eqnarray*}
 However, due to a clever re-summation of Hermite polynomials (which
we will not demonstrate) we have the much cleaner result

\begin{eqnarray}
\hat{f}_{j,n}^{\left(M\right)} & = & \sum_{k=0}^{n}{n \choose k}M^{n-k}\partial_{z}^{n-k}\left[\left(-N_{\phi}\right)^{k}\partial_{z}^{k}f_{j,0}\right],\label{eq:app_f_form}
\end{eqnarray}
where we have once again dropped $\frac{\left(2\i\right)^{n}}{M^{n}}\frac{\mathcal{N}_{n}}{\mathcal{N}_{0}}$
just as in the main text.

\section{Operators with periodic boundary conditions\label{sec:pbc-Operators}}

Similarly to the relation $e^{\i\pi\tau My^{2}}f_{j,n}^{\left(M\right)}=\phi_{j,n}^{\left(M\right)}$
in the main text, we may now define an operator equivalent of the
LLL projector $e^{\i\pi\tau N_{\phi}y^{2}}\hat{f}_{j,n}^{\left(M\right)}=\hat{\phi}_{j,n}^{\left(M\right)}e^{\i\pi\tau\left(N_{\phi}-M\right)y^{2}}$
for a general $n$:th Landau level. We may express $\hat{\phi}_{j,n}^{\left(M\right)}$
as a series expansion in 
\begin{equation}
\hat{g}_{n}=\left.\hat{f}_{n}^{\left(M\right)}\right|_{f_{0}\to\phi_{0}}=\left(M\partial_{z}-\tilde{\partial}_{z}N_{\phi}\right)^{n}\phi_{0},\label{eq:Goperator}
\end{equation}
 where we simply replace the $f_{0}$ in $\hat{f}_{j,n}^{\left(M\right)}$
by $\phi_{0}$. It is straight forward to show that the operator $\hat{g}_{n}$
satisfies the desired periodicity boundary boundary conditions 

\[
e^{\i2\pi N_{\phi}x}\tilde{t}\left(\tau L\right)\hat{g}_{n}=\hat{g}_{n}\tilde{t}\left(\tau L\right)e^{\i2\pi\left(N_{\phi}-M\right)x}
\]
 by repeating the arguments that where used in conjunction with eqn.
(\ref{eq:D_to_D}). The only difference is that now its an exponential
of $x$ and not $z$ that is considered. However, since $\left[\partial_{z},\left[\partial_{z},x\right]\right]=0=\left[\partial_{z},\left[\partial_{z},z\right]\right]$
the calculation is identical.

Considering now the function 
\[
\hat{\phi}_{n}^{\left(M\right)}=e^{\i\pi\tau N_{\phi}y^{2}}\hat{f}_{n}^{\left(M\right)}e^{-\i\pi\tau\left(N_{\phi}-M\right)y^{2}},
\]
 we can use (\ref{eq:Goperator}) to argue that $\hat{\phi}_{n}^{\left(M\right)}\neq\hat{g}_{n}$
but that there will also will be sub leading terms proportional to
$\hat{g}_{n-2},\hat{g}_{n-4},\ldots,\hat{g}_{0}$. Unlike the arguments
that where used in conjunction with eqn. (\ref{eq:D_to_D}) we are
now pulling exponentials of $y^{2}$ through $\hat{D}$, and since
$\left[\partial_{z},\left[\partial_{z},y^{2}\right]\right]\ne0$ the
shifts of $\partial_{z}$ and $\tilde{\partial}_{z}$ cannot be applied
independently. This is what leads to the sub leading terms. If we
define $\chi=M\left(N-M\right)N_{\phi}\frac{\pi\tau}{2\i\tau_{2}^{2}L^{2}}=\frac{M\left(N-M\right)\tau}{4\i\tau_{2}}$
then we may explicitly show that 
\begin{align*}
\hat{\phi}_{1} & =\hat{g}_{1}\\
\hat{\phi}_{2} & =\hat{g}_{2}+1\chi\hat{g}_{0}\\
\hat{\phi}_{3} & =\hat{g}_{3}+3\chi\hat{g}_{1}\\
\hat{\phi}_{4} & =\hat{g}_{4}+6\chi\hat{g}_{2}+3\chi^{2}\hat{g}_{0}\\
\hat{\phi}_{5} & =\hat{g}_{5}+10\chi\hat{g}_{3}+15\chi\hat{g}_{1}\\
\hat{\phi}_{6} & =\hat{g}_{6}+15\chi\hat{g}_{4}+45\chi^{2}\hat{g}_{2}+15\chi^{3}\hat{g}_{0}\\
\hat{\phi}_{7} & =\hat{g}_{7}+21\chi\hat{g}_{5}+105\chi^{2}\hat{g}_{3}+105\chi^{3}\hat{g}_{1}\\
\hat{\phi}_{8} & =\hat{g}_{8}+28\chi\hat{g}_{6}+210\chi^{2}\hat{g}_{4}+420\chi^{3}\hat{g}_{2}+105\chi^{4}\hat{g}_{0}.
\end{align*}
 This may be summarized as 
\begin{equation}
\hat{\phi}_{n}=\sum_{k=0}^{\left\lceil \frac{n}{2}\right\rceil }T\left(n,k\right)\chi^{k}\hat{g}_{n-2k},\label{eq:app_phi_expansion}
\end{equation}
 where $T\left(n,k\right)$ is the triangle of Bessel numbers (OEIS
series A100861)\cite{OEIS_Bessel}.
\end{document}